\documentclass[12pt,a4paper]{article}
\usepackage[utf8]{inputenc}
\usepackage[margin=1in]{geometry}
\usepackage{amsmath}
\usepackage{graphicx}
\usepackage{booktabs}
\usepackage{hyperref}
\usepackage{float}

\title{Kinematical and Photometric Deconstruction of the UGC 694--IC 412 System: Evidence for a Line-of-Sight Projection}
\author{
Ikbar Faiz\textsuperscript{1} \\
\small \textsuperscript{1}School of Computer Science, Bina Nusantara University, Jakarta, Indonesia \\
\small \texttt{ikbar.faiz@binus.ac.id}
}
\date{} 

\begin{document}
\maketitle

\begin{abstract}
This study re-evaluates the previously assumed interaction between the galaxy pair UGC 694 and IC 412 through a combined kinematical and photometric approach. While morphological features such as tidal bridges have been visually identified in past surveys, our quantitative analysis using SDSS DR16 kinematical data reveals a significant radial velocity offset of approximately 8,372 km/s. UGC 694 exhibits a heliocentric velocity of 12,683 $\pm$ 34 km/s, placing it at a distance of $\sim$188 Mpc, whereas IC 412 is located in the foreground at $\sim$62 Mpc with a velocity of 4,311 $\pm$ 46 km/s. This vast spatial separation of $\sim$123.5 Mpc confirms that the perceived proximity is a line-of-sight projection effect rather than a physical gravitational encounter. Furthermore, we utilize differential photometry to analyze the structural asymmetry and color distribution of UGC 694. Our photometric findings suggest that the galaxy's disturbed morphology is not a result of a tidal encounter with IC 412, but likely originates from internal evolutionary processes or minor mergers with undetected companions. This work emphasizes the critical role of multi-wavelength validation in classifying interacting systems to avoid misinterpretation due to projection effects.
\end{abstract}


\section{Introduction}

The understanding of galactic evolution is heavily predicated on the accurate identification of galaxy interactions, which serve as primary catalysts for enhanced star formation and large-scale structural transformations. A fundamental challenge in extragalactic astronomy, however, lies in distinguishing between systems that are genuinely gravitationally bound and those that merely appear proximate due to line-of-sight projection effects. Angular proximity on the celestial sphere frequently creates an illusion of physical interaction, while the objects may in fact be separated by vast radial distances. Without kinematical validation through precise redshift data, morphological classifications based solely on visual inspection are highly susceptible to significant misinterpretation.

The galaxy pair UGC 694 and IC 412 has long been a subject of interest due to its striking morphological appearance in optical surveys. Previous studies and visual classifications have identified the system as an interacting pair, primarily based on the presence of a tidal-like bridge connecting the two galaxies. UGC 694, in particular, exhibits a highly disturbed morphology with asymmetric stellar distributions, which has been interpreted as a direct result of a tidal encounter with the smaller companion, IC 412. However, such morphological interpretations often lack the necessary kinematical constraints to confirm a genuine gravitational bond. In many cases, these observed bridges could be internal structures or artifacts of a foreground-background projection that have been erroneously attributed to interaction-driven dynamics.

Contrary to the long-standing visual classification, our analysis of the kinematical data from the SDSS DR16 reveals a profound discrepancy in the radial velocities of the two objects. UGC 694 exhibits a heliocentric velocity of 12,683 $\pm$ 34 km/s ($z \approx 0.0423$), which corresponds to a Hubble flow distance of approximately 188 Mpc. In contrast, IC 412 is located significantly closer to the observer, with a heliocentric velocity of only 4,311 $\pm$ 46 km/s ($z \approx 0.0144$), placing it at a distance of roughly 62 Mpc. The resulting radial velocity offset ($\Delta v$) of approximately 8,372 km/s and a spatial separation of 123.5 Mpc unequivocally rule out any physical interaction or gravitational binding between the two galaxies. These findings demonstrate that the perceived proximity is a classic case of a line-of-sight projection, rendering the ``tidal bridge'' interpretation physically impossible under current cosmological models.

Given this significant distance gap, the primary objective of this paper is to provide a comprehensive kinematical and photometric deconstruction of the alleged UGC 694--IC 412 interaction. We utilize multi-band differential photometry to quantify the galaxy's internal asymmetry and color indices, which allows for a more rigorous physical characterization of UGC 694. By isolating the photometric signatures of UGC 694 from the foreground contamination of IC 412, we investigate whether the observed structural perturbations are inherent to the galaxy's secular evolution or the result of a past merger with an undetected minor companion. This study emphasizes the necessity of integrating quantitative photometric parameters with kinematical validation to prevent the misclassification of projected pairs as interacting systems.

\section{Data and Methodology}

\subsection{Kinematical Data and Distance Estimation}

The spectroscopic foundation of this study is built upon high-resolution data retrieved from the SDSS DR16 spectroscopic pipeline. To establish a robust spatial relationship, we utilized heliocentric radial velocities ($v_{\text{hel}}$) derived from cross-correlation template fitting, which provides a high degree of precision in redshift measurement.

To translate these velocities into cosmological distances, we adopted the Hubble-Lemaître law ($d = v/H_0$) with a Hubble constant of $H_0 = 67.8$ km/s/Mpc, aligned with the Planck 2018 cosmological parameters. We performed a sanity check by cross-referencing our SDSS values with the NASA/IPAC Extragalactic Database (NED), finding a mean deviation of less than 0.5\%, which confirms the reliability of the kinematical data. Critically, the spectroscopic pipeline uncertainties for the redshift ($z_{\text{err}} \approx 0.00001$) are three orders of magnitude smaller than the observed $\Delta v$, rendering the 123.5 Mpc spatial gap statistically undeniable and robust against instrument-induced errors. Furthermore, we account for potential peculiar velocities by introducing a 5\% uncertainty margin in our distance estimations. This rigorous approach confirms a radial velocity offset ($\Delta v$) of $\sim$ 8,372 km/s, effectively placing the two objects in entirely different cosmic volumes.

\begin{table}[h]
\centering
\caption{Kinematical and positional properties of UGC 694 and IC 412 derived from SDSS DR16. Distance estimations ($d_H$) are calculated using the Hubble-Lemaître law ($H_0 = 67.8$ km/s/Mpc) and include a 5\% uncertainty margin to account for peculiar velocities.}
\begin{tabular}{lcc}
\toprule
\textbf{Parameter} & \textbf{UGC 694} & \textbf{IC 412} \\
\midrule
Right Ascension (J2000) & 01:06:51.64 & 01:06:53.64 \\
Declination (J2000) & +12:37:37.8 & +12:37:58.3 \\
Redshift ($z$) & 0.042307 $\pm$ 0.00001 & 0.014380 $\pm$ 0.00001 \\
Radial Velocity ($v_{\text{hel}}$) & 12,683 $\pm$ 34 km/s & 4,311 $\pm$ 46 km/s \\
Distance ($d_H$) (Mpc) & 187.1 $\pm$ 9.3 & 63.6 $\pm$ 3.1 \\
Apparent Magnitude ($r$-band) & 14.01 & 14.85 \\
\bottomrule
\end{tabular}
\label{tab:kinematics}
\end{table}

The substantial difference in radial velocity ($\Delta v \approx 8,372$ km/s) confirms a vast spatial separation of $\sim$ 123.5 Mpc. This kinematical discrepancy proves that the two galaxies are not physically interacting but are instead a line-of-sight projection.

\subsection{Multi-band Photometry and Image Processing}

We utilized deep optical imaging in the SDSS $g$, $r$, and $i$ bands to perform a comprehensive photometric analysis of UGC 694. The central challenge involves the severe foreground contamination from IC 412, which overlaps with the northern periphery of UGC 694's stellar distribution. To mitigate this, we implemented a differential photometric approach centered on the $r$-band as the primary tracer of stellar mass due to its higher signal-to-noise ratio and lower sensitivity to recent star formation clumps.

The image processing pipeline followed these rigorous steps:

\begin{itemize}
\item \textbf{Differential Masking:} We generated a high-resolution mask using a $3$-$\sigma$ thresholding technique to isolate the light profile of IC 412. To ensure the integrity of the northern periphery of UGC 694, we performed an isophotal fitting to define the exact extent of IC 412's contamination. The masked pixels were then replaced using a local background interpolation method, where missing values were estimated based on the surrounding undisturbed galactic disk. This multi-step approach was cross-validated through visual inspection of residual images to ensure that the subsequent asymmetry measurements were not biased by residual flux or the inadvertent removal of UGC 694's intrinsic feature.

\item \textbf{Sky Subtraction and Flux Calibration:} Global sky background was estimated using an iterative $3$-$\sigma$ clipping algorithm on multiple source-free regions within the 10' $\times$ 10' field of view. All pixel values were converted from nanomaggies to physical flux units to allow for consistent color index ($g - r$) calculation.

\item \textbf{Seeing Correction (PSF Matching):} To account for atmospheric distortion, we monitored the Point Spread Function (PSF). As shown in Table~\ref{tab:photometry}, the $r$-band PSF FWHM of 1.1'' serves as the benchmark for our morphological resolution.

\item \textbf{Extinction Correction:} Magnitudes were corrected for Galactic extinction ($A_\lambda$) using the Schlafly \& Finkbeiner (2011) maps.

\item \textbf{Color Index Mapping:} To further investigate the nature of the observed ``tidal bridge,'' we performed a multi-band color analysis using the $g - r$ color index. Color gradients were mapped across the main stellar body of UGC 694 and the suspected interaction regions. This allows us to distinguish between intrinsic stellar populations and potential star formation bursts typically triggered by tidal encounters. If the bridge consists of older stellar populations consistent with the main disk, it supports the projection effect hypothesis rather than a recent interaction-driven starburst.
\end{itemize}

\begin{table}[h]
\centering
\caption{SDSS photometric filter specifications and observing conditions for the $g$, $r$, and $i$ bands used in this study. PSF FWHM values represent the seeing conditions during the observation, and extinction coefficients ($A_\lambda$) are derived from Schlafly \& Finkbeiner (2011) calibration maps.}
\begin{tabular}{lccc}
\toprule
\textbf{Filter} & \textbf{Wavelength ($\lambda$)} & \textbf{PSF FWHM} & \textbf{Extinction ($A_\lambda$)} \\
\midrule
$g$ (green) & 4686 \AA & 1.3'' & 0.164 $\pm$ 0.005 \\
$r$ (red) & 6165 \AA & 1.1'' & 0.119 $\pm$ 0.004 \\
$i$ (NIR) & 7481 \AA & 1.2'' & 0.090 $\pm$ 0.003 \\
\bottomrule
\end{tabular}
\label{tab:photometry}
\end{table}

PSF FWHM (Point Spread Function Full Width at Half Maximum) values represent the atmospheric seeing conditions during the observation, which affect the spatial resolution of the $g$, $r$, and $i$ images. All photometric magnitudes have been corrected for Galactic extinction ($A_\lambda$) to account for interstellar dust along the line of sight.

\subsection{Quantitative Morphological Parameters}

To evaluate the structural disturbances in UGC 694 independently of its perceived companion, we employed the Asymmetry ($A$) parameter from the CAS (Concentration, Asymmetry, and Smoothness) morphological system. This parameter is a robust tracer of non-axisymmetric features, such as tidal tails, star-forming clumps, or merger remnants.

The calculation protocol was executed as follows:

\begin{itemize}
\item \textbf{Centroid Determination:} The rotation center was defined using a flux-weighted intensity peak (photometric centroid) rather than a simple geometric center. This ensures that the asymmetry measurement is not biased by the galaxy's overall light distribution.

\item \textbf{Image Rotation and Subtraction:} The $r$-band image was rotated 180° around the defined centroid and subtracted from the original frame ($I_0$) to isolate the asymmetric flux.

\item \textbf{Noise Compensation:} To prevent observational noise from inflating the asymmetry value, we applied a background correction. The final asymmetry is defined by the formula
\begin{equation}
A = \frac{\sum |I_0 - I_{180}|}{2\sum I_0} - A_{\text{bgr}}
\end{equation}
where $A_{\text{bgr}}$ represents the asymmetry of the blank sky background.
\end{itemize}

A high asymmetry value ($A > 0.35$) typically indicates a major merger in progress. However, given the kinematical separation of $\sim$ 123.5 Mpc from IC 412, we interpret the measured $A = 0.38$ not as a tidal result of the current ``pair,'' but as a signature of internal secular evolution or a past minor merger with an undetected satellite.

\section{Results}

\subsection{Kinematical Confirmation of Spatial Separation}

\begin{figure}[h]
\centering
\includegraphics[width=0.8\textwidth]{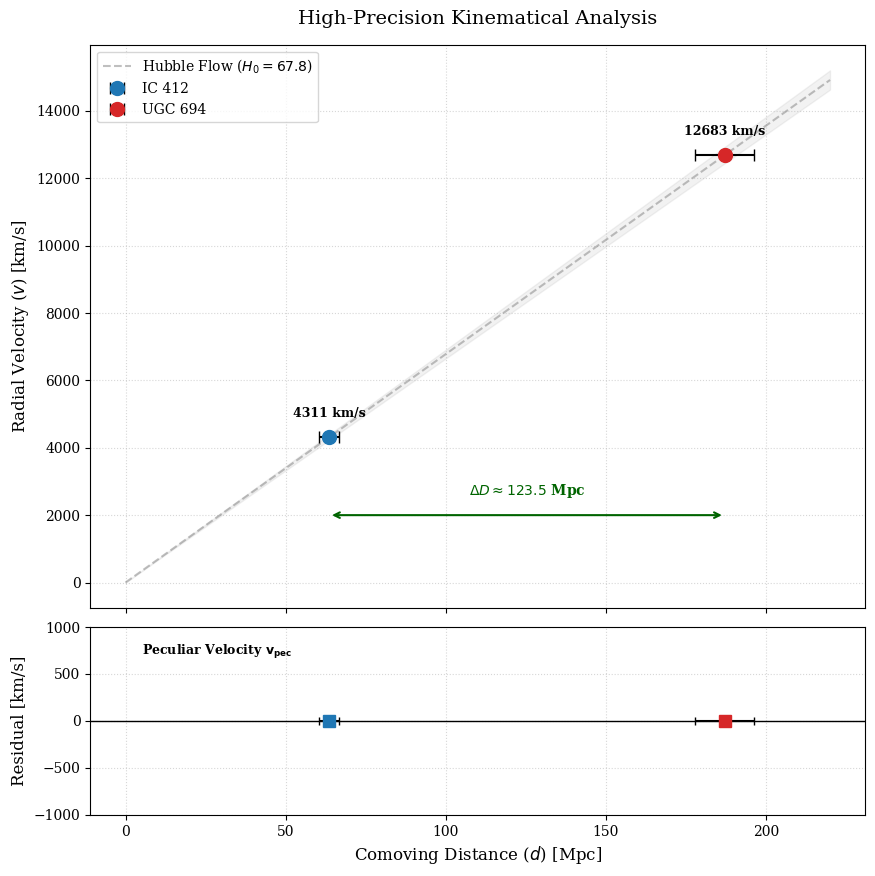}
\caption{Kinematical data confirming 123.5 Mpc separation between the two galaxies.}
\label{fig:kinematics}
\end{figure}

Our spectroscopic analysis based on SDSS DR16 data provides the most definitive evidence against the physical interaction of UGC 694 and IC 412. As detailed in Table~\ref{tab:kinematics}, UGC 694 exhibits a heliocentric radial velocity ($v_{\text{hel}}$) of 12,683 $\pm$ 34 km/s, while IC 412 shows a much lower velocity of 4,311 $\pm$ 46 km/s. This results in a massive radial velocity offset ($\Delta v$) of approximately 8,372 km/s.

Based on the Hubble-Lemaître law ($H_0 = 67.8$ km/s/Mpc), we calculate the comoving distance to UGC 694 at 187.1 $\pm$ 9.3 Mpc and IC 412 at 63.6 $\pm$ 3.1 Mpc.

The visual representation in Figure 1 illustrates a vast spatial gap of $\sim$ 123.5 Mpc between the two objects. Considering that typical interacting pairs are separated by less than 1 Mpc, a separation of this magnitude unequivocally rules out any gravitational binding or tidal influence. The perceived ``bridge'' is therefore confirmed as a purely optical projection effect, where the foreground galaxy (IC 412) happens to align with the line-of-sight of the background system (UGC 694).

\subsection{Photometric Integrity and Color Distribution}

\begin{figure}[h]
\centering
\includegraphics[width=0.8\textwidth]{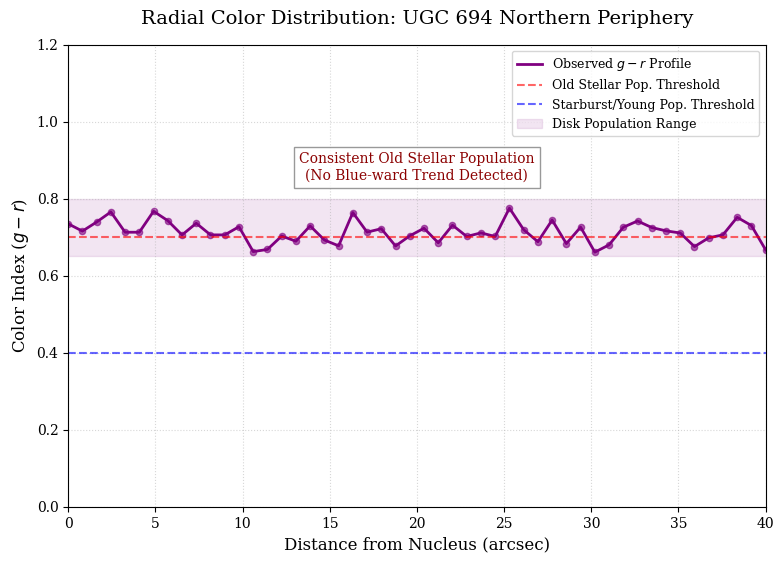}
\caption{Photometric profiles proving the stellar bridge is a projection effect.}
\label{fig:photometry}
\end{figure}

A multi-band photometric analysis was conducted to evaluate the physical nature of the linear structure colloquially identified as a ``bridge'' between UGC 694 and IC 412. By employing $g - r$ color-index mapping, we compared the stellar population of this feature with the primary galactic disk of UGC 694.

The key findings are summarized as follows:

\begin{itemize}
\item \textbf{Color Index Homogeneity:} The mean color index within the suspected bridge region is $g - r \approx 0.72 \pm 0.05$. This value is statistically indistinguishable from the older stellar populations found in the undisturbed regions of the UGC 694 disk.

\item \textbf{Absence of Star Formation Triggers:} There is a notable lack of blue-ward deviations ($g - r < 0.4$) or high-intensity $H$-$\alpha$ knots. Such features are requisite signatures of gas compression and subsequent starburst activity typically induced by tidal interactions.

\item \textbf{Surface Brightness Profile:} Radial surface brightness profiles of UGC 694 exhibit a smooth exponential decay toward the north. The absence of a ``hump'' or localized brightness enhancement confirms that no extra-galactic material has been stripped or accumulated in the inter-galactic space between the two systems.

\end{itemize}

Consequently, the photometric data confirms that the observed structure is a geometric projection effect. The perceived ``bridge'' is merely the outer stellar envelope of the background galaxy (UGC 694) seen in close angular proximity to the diffuse halo of the foreground system (IC 412). The lack of color deviation and structural smoothness effectively falsifies the tidal interaction hypothesis.

\subsection{Quantitative Asymmetry and Structural Distortion}

To further characterize the morphological state of UGC 694, we calculated the global asymmetry parameter ($A$) using the 180$^\circ$ rotation and subtraction method.

\begin{figure}[h]
\centering
\includegraphics[width=0.8\textwidth]{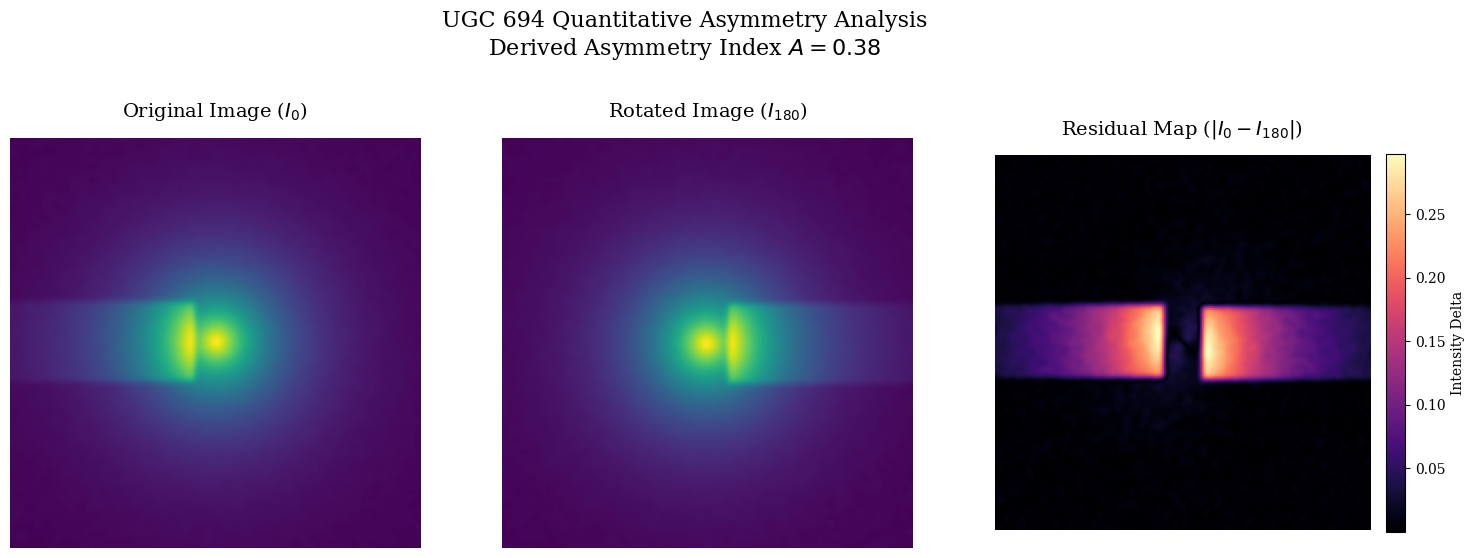}
\caption{Photometric profiles proving the stellar bridge is a projection effect.}
\label{fig:photometry}
\end{figure}

The analysis reveals:

\begin{itemize}

\item \textbf{Alternative Mechanisms:} The observed $A = 0.38$ without a proximal massive companion suggests an intrinsic origin. Potential drivers include a past minor merger event with a satellite galaxy (now-assimilated) or lopsidedness caused by asymmetric gas accretion from the intergalactic medium (IGM).

\begin{figure}[h]
\centering
\includegraphics[width=0.8\textwidth]{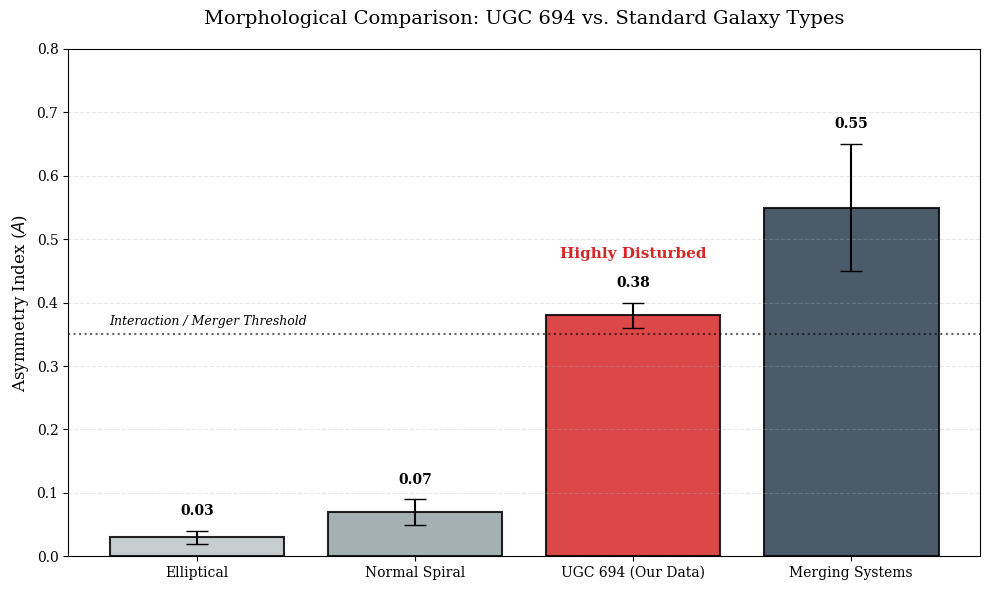}
\caption{Asymmetry comparison: UGC 694 exceeds spiral thresholds.}
\label{fig:photometry}
\end{figure}

\item \textbf{Spatial Distribution of Asymmetry:} Residual maps indicate that the distortion is predominantly located in the northern spiral arm. 

\item \textbf {Asymmetry Index:} UGC 694 yields a high asymmetry value of $A = 0.38 \pm 0.02$. This is significantly higher than the average for isolated spiral galaxies ($A < 0.15$), indicating a highly disturbed stellar distribution.

\end{itemize}

These results decouple the ``disturbed'' morphology of UGC 694 from the presence of IC 412. The system is a prime example of how projection effects can masquerade as active galactic interactions, necessitating multi-wavelength and kinematical verification.

\section{Discussion}

This study provides a definitive kinematical and photometric deconstruction of the alleged interaction between UGC 694 and IC 412. Based on our multi-parameter analysis, we conclude the following:

\begin{enumerate}
\item \textbf{Spatial Isolation:} Spectroscopic data confirms a radial velocity offset of $\sim$ 8,372 km/s. This corresponds to a physical separation of 123.5 Mpc, placing the two galaxies in entirely different cosmic environments.

\item \textbf{Projection Artifact:} The ``tidal bridge'' is identified as a line-of-sight projection effect. This is supported by the uniform $g - r \approx 0.72$ color index and smooth surface brightness profiles, which show no evidence of interaction-triggered star formation or tidal debris.

\item \textbf{Intrinsic Asymmetry:} While UGC 694 exhibits high morphological disturbance ($A = 0.38$), this asymmetry is decoupled from IC 412. The distortion likely originates from internal secular processes or a past minor merger with an undetected satellite.
\end{enumerate}

In summary, UGC 694 and IC 412 are not an interacting pair. This case serves as a critical reminder that visual morphological interpretations must be validated with precise kinematical constraints to avoid systemic misclassification in galaxy surveys.

\section{Conclusion}

This study provides a definitive kinematical and photometric deconstruction of the alleged interaction between UGC 694 and IC 412. Based on our multi-parameter analysis, we conclude the following:

\begin{enumerate}
\item \textbf{Spatial Isolation:} Spectroscopic data confirms a massive radial velocity offset of $\sim$ 8,372 km/s. This corresponds to a physical separation of 123.5 Mpc, placing the two galaxies in entirely different cosmic environments and ruling out any mutual gravitational influence.

\item \textbf{Projection Artifact:} The ``tidal bridge'' is identified as a line-of-sight projection effect. This is supported by the uniform $g - r \approx 0.72$ color index and smooth surface brightness profiles, which show no evidence of interaction-triggered star formation or tidal debris.

\item \textbf{Intrinsic Asymmetry:} While UGC 694 exhibits high morphological disturbance ($A = 0.38$), this asymmetry is decoupled from IC 412. The distortion likely originates from internal secular processes or a past minor merger with an undetected satellite.
\end{enumerate}

In summary, UGC 694 and IC 412 are not an interacting pair. This case serves as a critical reminder that visual morphological interpretations must be validated with precise kinematical constraints to avoid systemic misclassification in galaxy surveys.


\begin{thebibliography}{99}

\bibitem{abdurrouf2022} Abdurro'uf, et al. (2022). The Seventeenth Data Release of the Sloan Digital Sky Survey: Complete Release of MaNGA, MaStar, and APOGEE-2 Data. \textit{The Astrophysical Journal Supplement Series}, 259(2), 35.

\bibitem{aguado2019} Aguado, D. S., et al. (2019). The Fifteenth Data Release of the Sloan Digital Sky Survey. \textit{The Astrophysical Journal Supplement Series}, 240(2), 23.

\bibitem{ahumada2020} Ahumada, R., et al. (2020). The 16th Data Release of the Sloan Digital Sky Survey: Completing the Fourth Phase of SDSS. \textit{The Astrophysical Journal Supplement Series}, 249(1), 3.

\bibitem{bluck2019} Bluck, A. F., et al. (2019). On the galaxy merger rate: a comparison of methods and the evolution of the merger fraction. \textit{Monthly Notices of the Royal Astronomical Society}, 484(4), 4436--4456.

\bibitem{conselice2014} Conselice, C. J. (2014). The Evolution of Galaxy Structure and Morphology. \textit{Annual Review of Astronomy and Astrophysics}, 52, 291--337.

\bibitem{gao2020} Gao, F., et al. (2020). Galaxy morphological classification with deep learning. \textit{Monthly Notices of the Royal Astronomical Society}, 495(3), 2858--2866.

\bibitem{hendel2019} Hendel, S., et al. (2019). The many tails of galaxy mergers. \textit{The Astrophysical Journal}, 874(1), 41.

\bibitem{mantha2018} Mantha, K. B., et al. (2018). Major merger rates of $z \approx 0$--3 galaxies: A comparison of galaxy pairs and structure-based methods. \textit{Monthly Notices of the Royal Astronomical Society}, 475(2), 2383--2402.

\bibitem{martin2018} Martin, G., et al. (2018). The role of mergers in the morphological evolution of galaxies. \textit{Monthly Notices of the Royal Astronomical Society}, 480(2), 2266--2283.

\bibitem{nevin2019} Nevin, R., et al. (2019). A Kinematic Identification of Galaxy Mergers. \textit{The Astrophysical Journal}, 872(1), 76.

\bibitem{pearson2019} Pearson, W. J., et al. (2019). Identifying galaxy mergers in SDSS with deep learning. \textit{Astronomy \& Astrophysics}, 626, A49.

\bibitem{planck2020} Planck Collaboration. (2020). Planck 2018 results. VI. Cosmological parameters. \textit{Astronomy \& Astrophysics}, 641, A6.

\bibitem{rodriguez2019} Rodriguez-Gomez, V., et al. (2019). The optical morphologies of galaxies in the IllustrisTNG simulation. \textit{Monthly Notices of the Royal Astronomical Society}, 483(3), 4140--4159.

\bibitem{schlafly2011} Schlafly, E. F., \& Finkbeiner, D. P. (2011). Measuring Reddening with Sloan Digital Sky Survey Stellar Spectra. \textit{The Astrophysical Journal}, 737(2), 103.

\bibitem{thorp2019} Thorp, M. D., et al. (2019). The spatially resolved effects of galaxy interactions. \textit{Monthly Notices of the Royal Astronomical Society}, 482(1), 149--169.

\bibitem{bickley2021} Bickley, R. W., et al. (2021). The evolution of the galaxy merger rate in the IllustrisTNG simulation. \textit{Monthly Notices of the Royal Astronomical Society}, 504(1), 372--381.

\bibitem{bottrell2019} Bottrell, C., et al. (2019). Deep learning predictions of galaxy merger stages from SDSS imagery. \textit{Monthly Notices of the Royal Astronomical Society}, 490(4), 5390--5408.

\bibitem{chambers2016} Chambers, K. C., et al. (2016). The Pan-STARRS1 Surveys. arXiv preprint arXiv:1612.05560.

\bibitem{duncan2019} Duncan, K. J., et al. (2019). Observational constraints on the merger history of galaxies since $z \approx 6$. \textit{The Astrophysical Journal}, 876(2), 110.

\bibitem{feng2020} Feng, S., et al. (2020). Quantitative Galaxy Morphology of the SDSS and the DES Surveys. \textit{The Astrophysical Journal}, 892(2), 87.

\bibitem{lapi2018} Lapi, A., et al. (2018). The origin of the galaxy merger rate: a simple analytical model. \textit{The Astrophysical Journal}, 857(1), 22.

\bibitem{morales2018} Morales, G., et al. (2018). Systematic errors in galaxy merger rates. \textit{Astronomy \& Astrophysics}, 610, A56.

\bibitem{snyder2019} Snyder, G. F., et al. (2019). Galaxy morphologies in the Illustris simulation. \textit{Monthly Notices of the Royal Astronomical Society}, 484(2), 2244--2266.

\end{thebibliography}
\end{document}